\documentclass[sigconf]{acmart}

\AtBeginDocument{%
  \providecommand\BibTeX{{%
    \normalfont B\kern-0.5em{\scshape i\kern-0.25em b}\kern-0.8em\TeX}}}

\copyrightyear{2020}
\acmYear{2020}
\setcopyright{usgovmixed}\acmConference[HotNets '20]{Proceedings of the 19th ACM
Workshop on Hot Topics in Networks}{November 4--6, 2020}{Virtual Event, USA}
\acmBooktitle{Proceedings of the 19th ACM Workshop on Hot Topics in Networks
(HotNets '20), November 4--6, 2020, Virtual Event, USA}
\acmPrice{15.00}
\acmDOI{10.1145/3422604.3425924}
\acmISBN{978-1-4503-8145-1/20/11}
\newcommand{\parahead}[1]{\vspace{2pt plus 1pt minus 1pt}\noindent{\bfseries #1}}
\newcommand{\parabreak}{\vspace*{1.00ex minus 0.25ex}\noindent}

\usepackage[utf8]{inputenc}
\usepackage[english]{babel}
\usepackage{graphicx,booktabs,physics,amsmath,times}
\usepackage{tabularx,tablefootnote,wrapfig,multirow}
\usepackage{hyphenat,xspace,color,colortbl,enumitem}
\usepackage[labelformat=simple,skip=0pt]{subcaption}
\begin{document}

\fancyhead{}
\title{Towards Hybrid Classical-Quantum Computation Structures in Wirelessly-Networked Systems}


\author{Minsung Kim}
\affiliation{%
 \institution{Princeton University}
 \institution{NASA Ames Research Center, QuAIL}
 \institution{USRA Research Institute for Advanced Computer Science}
}

\author{Davide Venturelli}
\affiliation{%
 \institution{NASA Ames Research Center, QuAIL}
 \institution{USRA Research Institute for Advanced Computer Science}
}

\author{Kyle Jamieson}
\affiliation{%
 \institution{Princeton University}
}
\renewcommand{\shortauthors}{Trovato and Tobin, et al.}

\begin{abstract}
  With unprecedented increases in traffic load in today's wireless networks, design challenges shift from the wireless network itself to the computational support behind the wireless network. In this vein, there is new interest in quantum-computing approaches because of their potential to substantially speed up processing, and so improve network throughput. However, quantum hardware that actually exists today is much more susceptible to computational errors than silicon-based hardware, due to the physical phenomena of decoherence and noise.  This paper explores the boundary between the two types of computation---classical-quantum hybrid processing for optimization problems in wireless systems---envisioning how wireless can simultaneously leverage the benefit of both approaches. We explore the feasibility of a hybrid system with a real hardware prototype using one of the most advanced experimentally available techniques today, \emph{reverse quantum annealing.} Preliminary results on a low-latency, large MIMO system envisioned in the 5G New Radio roadmap are encouraging, showing approximately 2--10$\times$ better performance in terms of processing time than prior published results.

\end{abstract}

\ccsdesc[500]{Networks~Wireless access points, base stations and infrastructure}
\ccsdesc[500]{Computer systems organization~Quantum computing}



\keywords{Wireless Networks, Quantum Computing, Hybrid Design, Quantum Annealing, MIMO Detection}


\maketitle

\section{Introduction}

The demand for faster and more capable video, audio, 
and teleconferencing applications over the 
past decade has resulted in sharp increases of wireless traffic 
loads at base stations.
To increase the \emph{spectral efficiency} of the entire network, 
there are many techniques available, most notably the 5G suite of technologies,
including Massive MIMO, dense small cells, and millimeter wave communication.
These techniques can indeed increase spectral efficiency to a point,
but to scale up in terms of users and traffic loads, they quickly demand 
exponentially 
more \emph{computational} throughput, at low \emph{computational}
latencies \cite{andrews2014will}.
The latter challenge arises from the 
limited available processing time before a radio must turn\hyp{}around response to incoming data,
a part of the link layer's automatic repeat request functionality
in almost all wireless networked systems,
such as wireless local area and 4G LTE cellular networks \cite{BigStation, dahlman20134g,dot11ac-spec,dot11-2012}. 
Forthcoming ultra\hyp{}low latency designs,
envisioned in 5G \emph{New Radio} and
beyond \cite{5g-jsac17}, tighten these requirements even further.

\begin{figure}
\centering
\includegraphics[width=0.9\linewidth]{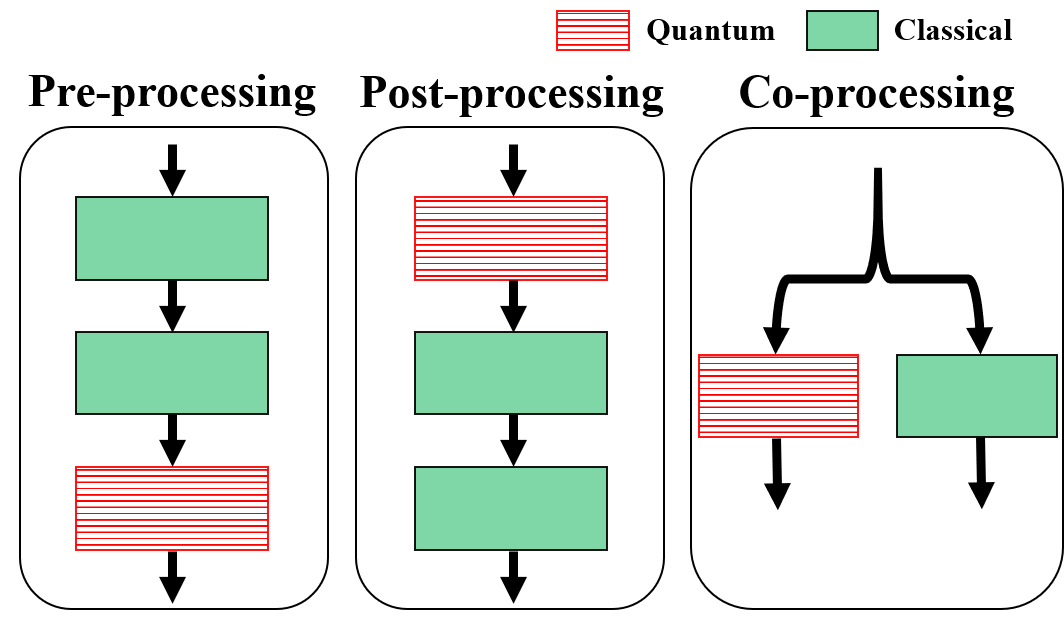}
\caption{Fundamental computation structures for the coordination 
of classical\hyp{}quantum processing units.}
\label{f:highleveldesign}
\end{figure}

There has been much recent effort dedicated 
to faster and more parallel integration of 
silicon processors into individual chip packages. 
Indeed, parallel computing is mainstream for 
high-performance computation at wireless 
base stations \cite{mathar2001optimal, baker2000cluster, 
zeng1998glomosim}. However, rental of physical space
and physical cooling dominate the cost of operating a 
base station 
and so the consensus architecture in cellular wireless networks is
based on a principle of aggregating many base stations' processing
together in a centralized data center via 
low\hyp{}latency optic fiber, an architecture called a
\emph{centralized radio access network} (C-RAN) \cite{checko2014cloud, lin2010wireless, sundaresan2013cloud}. However, 
even in a C-RAN architecture, practical constraints 
(in algorithms of wireless networks and\fshyp{}or 
computer architectures) continue to limit 
speedup gains on classical computing platforms \cite{flexcore-nsdi17,king2019quantum,hill2008amdahl}.

\parabreak{}This paper takes a long view on these issues:\ we 
hypothesize that 
the aforementioned growth trends in wireless network demand 
will continue, and 
so in one to two decades, silicon\hyp{}based computational 
resources will likely not suffice, since their clock speed has 
already reached a plateau~\cite{courtland}, and routing, power, and 
heat will constrain parallelism in the future.

To overcome these fundamentals, researchers are beginning to 
explore quantum computing (or quantum-inspired algorithms~\cite{swendsen1986replica, metropolis1949monte, georgii2011gibbs})
as an alternative to classical computers to solve intractable optimization 
problems in wireless systems~\cite{kim2019leveraging, gao2009quantum, botsinis2014fixed, wang2016quantum}.
While error\hyp{}corrected 
quantum computation may eventually enable asymptotic speedups over 
the best possible classical algorithms \cite{preskill2018quantum},
at its current level of maturity, 
universal large-scale quantum computers 
are still many years away and so
quantum computers are termed
\emph{Noisy Intermediate\hyp{}Scale Quantum} (NISQ) devices.
The power of near\hyp{}term quantum algorithms~\cite{farhi2014quantum, Hadfield:2017:QAO:3149526.3149530,AQC, QA}, which are heuristic 
methods that tolerate relatively high levels of noise,
is currently unknown and the wide belief
is that it should be investigated empirically, due to the 
impossibility of simulating the hardware accurately. 
Despite the lack of theoretical guidance, the payoff of 
understanding these methods' performance in 
these early days could be transformative, similar to what has recently
happened in the field of deep learning,
thanks to the appearance of computational devices 
that allowed a sufficiently large models to be 
trained and tested in practice~\cite{lecun2015deep}.



The overarching goal of this paper is 
to give a preliminary assessment of the viability of hybrid 
classical\hyp{}quantum computational structures (as shown in 
Figure~\ref{f:highleveldesign}) on NISQ computers for wireless networks, 
and to stimulate novel designs of quantum\hyp{}enabled base stations
(in cellular networks) or access points (in Wi-Fi local\hyp{}area 
networks) in light of expected improvements in engineering 
and systems integration of this nascent technology in the 
next few years. Our two contributions are as follows:

\parabreak{}\textbf{1)} We outline a vision of classical\hyp{}quantum hybrid processing in wireless networked systems that aims to take advantage of both classical and quantum processing. For this paradigm, we share lessons that we learn from possible designs. For example, we explore the feasibility of a classical\hyp{}quantum hybrid algorithm enabled by a new quantum technique on an analog quantum annealing device.   

\parabreak{}\textbf{2)} On the quantum annealing device we compare the 
performance of a \emph{Reverse Annealing} (RA) classical\hyp{}quantum hybrid protocol 
against two fully quantum solvers: \emph{Forward Annealing} (FA) and a 
newly developed \emph{Forward\hyp{}Reverse} (FR) annealing. To our knowledge
this is the first time reverse annealing techniques have been attempted on wireless applications. 


While many computationally heavy problems exist that cause bottlenecks in 5G 
wireless networked systems (such as large-scale channel coding, resource 
allocation, scheduling, and pre\hyp{}coding), we pick \emph{Large MIMO 
detection}, an essential technique to increase wireless 
throughput that enables parallel data streams for many users 
(\emph{spatial multiplexing}) in the same wireless spectrum. However, to make full 
use of spatial multiplexing, much more sophisticated receiver 
designs with (near) optimal detectors are required~\cite{BigStation, argos-mobicom12,
flexcore-nsdi17, Geosphere}.  Preliminary work has shown the possibility of 
reducing these problems to a form amenable to solution on a quantum computer, and 
provides performance baselines 
\cite{botsinis2013quantum,kim2019leveraging,gao2009quantum, botsinis2014fixed}.

\parabreak{}Our experimental results show that RA starting from a candidate solution obtained by a fast greedy search outperforms all other tested methods for all channel modulations in terms of required compute time to finding the optimal decoding. 
For reference, for an eight\hyp{}user, 16-QAM detection/decoding problem, our version of RA achieves approximately up to 10$\times$ higher success probability than the previously published results for FA. The application\hyp{}specific classical solvers and the compilation parameters are standard and have not been tailored, so this result can be improved further without any hardware advances. In this regard, we discuss possible design thoughts towards our eventual design.

\section{Background and Related Work}
\label{s:backgorund}


\parahead{Quantum computers for optimization:} Quantum computers feature hardware that uses unique information processing capabilities based on quantum mechanics to perform calculations. While there remain much to investigate on the ultimate power of quantum computing, optimization is one of the key applications that the quantum computing community has 
identified as interesting in the short\hyp{}term due to two main approaches that have shown promise 
in NISQ devices:\ \emph{Quantum Annealing} (QA)~\cite{AQC, QA} and \emph{Quantum Approximate 
Optimization Algorithms} (QAOA)~\cite{farhi2014quantum, Hadfield:2017:QAO:3149526.3149530}.  
While QA and QAOA
are different hardware (the former is analog, the latter 
digital) they have in common that both methods work on classical combinatorial problems 
which are often expressed as an \emph{Ising Model}. For the purpose of this work, this model is trivially equivalent to a \emph{Quadratic 
Unconstrained Binary Optimization} (QUBO) form \cite{boros2007local, QUBO}, which can be expressed as finding the bitstring $\hat{q}_1,\ldots,\hat{q}_N$ that minimizes the cost function:
\begin{align}
E(\left\{ q_1, \ldots, q_N \right\})=
 \sum^{N_v}_{i\leq j} Q_{ij}q_i q_j,
\label{eqn:qubo}
\end{align}

where $N_v$ is variable count and $\mathbf{Q} \in \mathbb{R}^{N
\times N}$ is upper triangular matrix, each of elements representing QUBO coefficients and each of binary variables $q$ either 0 or 1. Quantum-inspired computing devices are often called \emph{Ising Machines}~(see e.g. \cite{inagaki2016coherent}), implementing hardware that physically encodes the objective function \ref{eqn:qubo} as a measurable observable of a physical system.  The workflow of both quantum and quantum-inspired methods can be usually described as follows: After reducing the problem of interest into the Ising/QUBO form and programming the form on the quantum hardware, \emph{Quantum Processing Unit} (QPU) on the hardware solves it in a way typical of  heuristics blackboxes.\footnote{For details on how it works in some concrete use cases, see for instance Refs~\cite{boixo2014evidence,denchev-44814,AQC, QA,farhi2014quantum, Hadfield:2017:QAO:3149526.3149530}.} In general, multiple calls $N_s$ are made to the device and the best sample (e.g. the one with the lowest QUBO cost function) is selected as the final solution.


\parahead{Classical-quantum hybrid approaches:} Since quantum computers were realized in experimental hardware, there have already been some studies on classical-quantum hybrid approach despite its relative novelty.
For instance, the sequential use of a classical method as input for a quantum one (or viceversa) is a straightforward hybrid approach. In \cite{tran2016hybrid}, quantum annealing was used to generate solution candidates checked by classical computing that more easily deal with hard constraints. Classical computing can also ease the problem by prefixing some variables as part of iterative loops~\cite{karimi2017boosting} or as a pre-processing~\cite{glover2018logical,qubo-preproc,lewis2017quadratic}. Other examples of hybridization include locality reduction simplifications~\cite{binaryreduction, ishikawa2014higher}, problem decompositions~\cite{zintchenko2015local, rosenberg2016building}. A solver block design consisting of multiple quantum annealing processors hybridized with Tabu search is also in the commercial offerings of D-Wave Systems~\cite{DwaveHybrid}. A qualitatively novel hybrid scheme introduced recently is known as reverse annealing~\cite{ReverseVenturelli, terry2020quantum}, which we further explore with our initial prototype in Section~\ref{s:exploratory} for MIMO detection problems.

\section{Design Challenges}
\label{s:design}


\begin{figure}
\centering
\includegraphics[width=\linewidth]{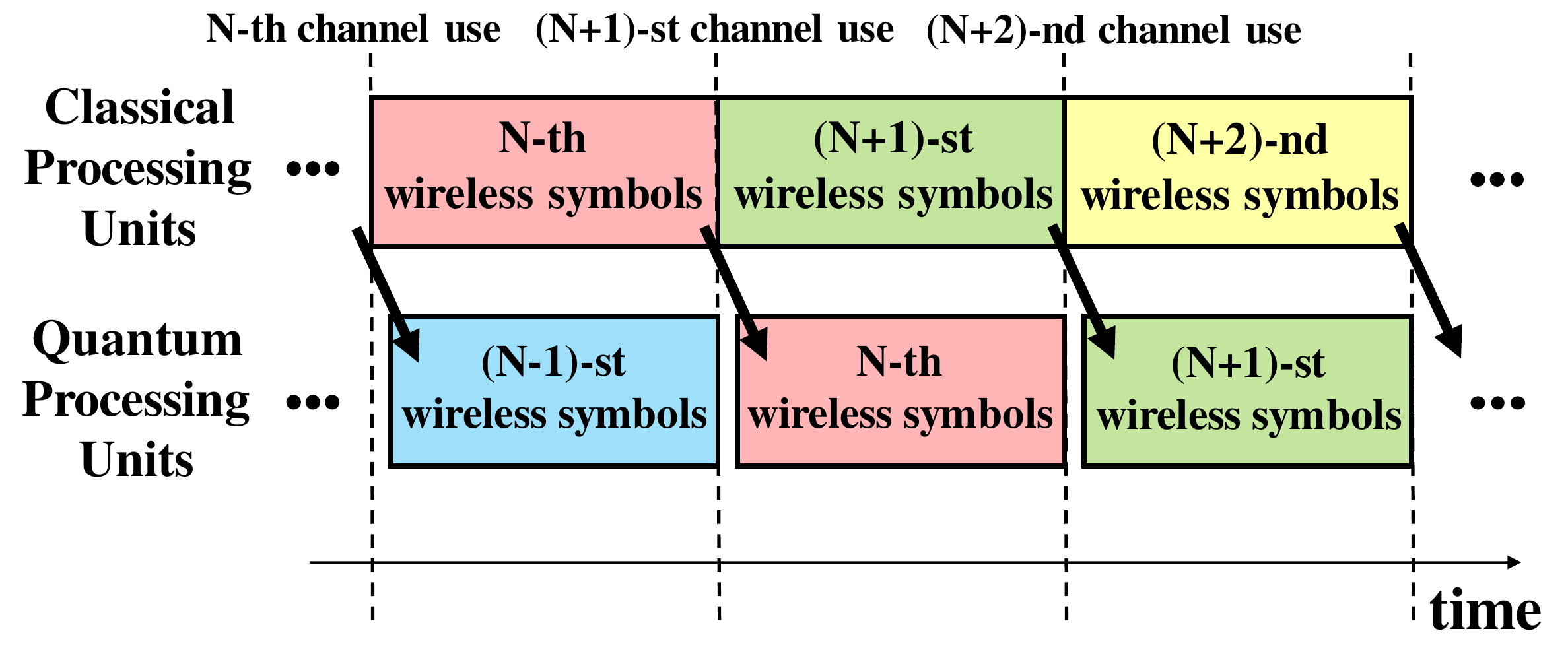}
\caption{Example pipeline design of hybrid computational structure for successive wireless channel uses.}
\label{f:pipeline}
\end{figure}

This section introduces three design challenges we faced in our exploration, as well as early design attempts that failed to deliver significant advantage by means of hybridization, for illustrative and educational purposes. 

\parahead{Challenge 1: Hybridization design.} The most crucial 
challenge consists in choosing the partitioning scheme of the computation between 
classical and quantum processing units. Pre-processing, post-processing and co-processing orchestration between quantum and classical modules need to be well specified.
To our best knowledge, none of classical-quantum hybrid techniques have been empirically tested on problems in wirelessly networked systems, on a real hardware prototype.

\parahead{Challenge 2: Optimal parameters.} The designed hybrid system will
include specifications on parameters crucial to
the performance of the quantum and classical optimizers. The best (interdependent) parameter values for each processing unit, \emph{i.e.}, those that maximize 
end\hyp{}to\hyp{}end performance of the system,
need to be identified.

\parahead{Challenge 3: Pipelining classical and quantum computation.}
The eventual goal of a hybrid classical\hyp{}quantum 
architecture is to enable \emph{ultra-high throughput} 
networks. To this end, how to best 
divide sequential computational tasks into classical \emph{or} quantum units 
and then assign those units to staged processing units is crucial. 
Fortunately, the nature of the sequential arrival of traffic over a
wireless link lends itself to such pipelining, as illustrated
in Figure~\ref{f:pipeline} where data bits from successive ``channel uses''
are processed in stages of the computational pipeline. 
However, typically more complex considerations are required for a pipelined system such as balancing, buffering, and costs and these become even more challenging in a classical-quantum hybrid system. 

\subsection{Initial Attempts}
\label{s:attemps}

In this section, we introduce our two initial attempts of hybrid algorithms for wireless applications that achieve no or very limited gain with respect to the non-hybrid use of the quantum machine. However, note that while these schemes were not beneficial in our prototype, they may work with more sophisticated algorithms and/or with next-generation quantum devices. 

\begin{figure}
\centering
\includegraphics[width=0.95\linewidth]{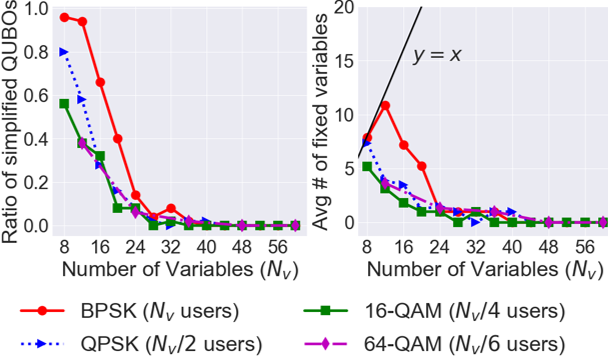}
\caption{Empirical results of the simplifying QUBO scheme for 50 instances of MIMO detection across different problem sizes and modulations: \emph{(Left)} ratio of simplified QUBOs and \emph{(Right)} average number of fixed variables in the simplified cases. The scheme achieves nearly no-effect for problems over 32-40 variables.}
\label{f:preprocessing}
\end{figure}

\parahead{Simplifying the QUBO form:}
A QUBO may be able to be simplified through the execution of a simple classical algorithm that would determine the value of some binary variables before quantum processing, depending on the value of its coefficients~\cite{lewis2017quadratic}. For $Q_{ii}>0$ ($\forall i$), if the absolute sum of negative coefficients of ($\forall k$) $Q_{ik}$ and $Q_{ki}$ ($<0$ and $k\neq i$) is less than $Q_{ii}$, then the i-th variable can be fixed to 1. In a similar way, certain QUBO variables can be prefixed into 0. Prefixing one logical variable reduces the search space by a factor of two. We test this scheme varying the problem size and modulations in Figure~\ref{f:preprocessing}. We observe that no simplifications are detected when the MIMO problems have over 32--40 variables, regardless of the chosen modulations. Given that computationally\hyp{}heavy optimization problems in envisioned 5G wireless networks contain more than 50 or even hundreds of variables, this pre\hyp{}processing scheme does not seem particularly useful.   

\begin{figure}
\centering
\includegraphics[width=0.80\linewidth]{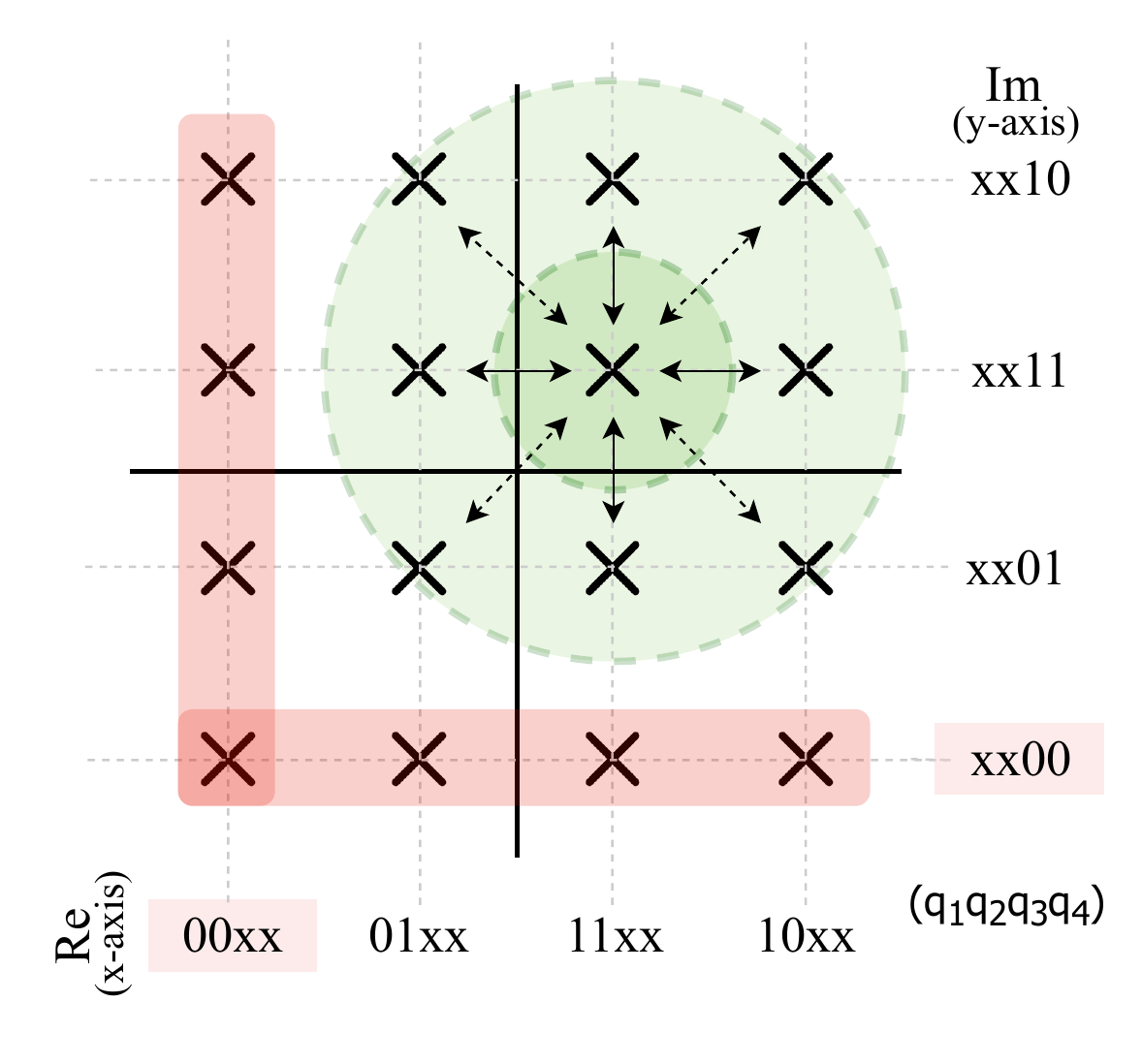}
\caption{An example of adding constraints based on a pre-knowledge of symbols on a gray-coded constellation of 16-QAM modulation. Given that $q_1q_2q_3q_4$ is very likely to be close to 1111, adding constraint terms $C_1\cdot(q_1-1)
\cdot(q_2-1)$ and $C_2\cdot(q_3-1)\cdot(q_4-1)$ to the QUBO forces the searching process to avoid unlikely symbols (red color-coded).} 
\label{f:constraints}
\end{figure}

\parahead{Soft information to narrow the search space:} Pre\hyp{}knowledge of variables (wireless symbols) that are very likely to be (un)assigned a certain value in the unknown global optimum solution might ease the problem. This pre\hyp{}knowledge is equivalent to soft\hyp{}information in wireless networks which are already applied in many applications and can be obtained in various ways~\cite{flexcore-nsdi17,larsson2008fixed, karimi2017boosting, zheng2006ldpc,tse-viswanath}. We test a new conservative algorithm of utilizing the pre-knowledge, by adding constraints as a classical pre-processing to force following quantum processing to search only spaces of more promising QUBO variables or symbols. For instance in Figure~\ref{f:constraints}, assuming we learn highly possible spaces (green color-coded), then
we can add constraint terms to the original QUBO to avoid unlikely spaces (red color-coded) without harming the global optimum (ideally). This approach seemingly looks useful, but it is difficult to find proper constraint factors on noisy, analog quantum machines especially for problems of more variables and multiple constraint factors and our empirical investigations have shown that it is not currently practical. 
\section{An Exploratory Study}
\label{s:exploratory}

In this section, we present an early-stage framework of classical\hyp{}quantum processing for a specific problem (MIMO detection) and method. We introduce our initial prototype design in Sec~\ref{s:prototype_design} and its implementation on the analog-based quantum device in Sec~\ref{s:implementation}. Results are in Sec~\ref{s:evaluation}.

\subsection{Prototype Design}
\label{s:prototype_design}
Our hybrid prototype design consists of two sequential modules; a Greedy Search (GS) classical algorithm and then a Reverse Annealing (RA) algorithm implemented on a \emph{D-Wave 2000Q} quantum annealer. GS is a very simple deterministic QUBO solver featuring linear complexity. RA is a variation of quantum annealing (see Figure~\ref{f:annealing}) where computation starts from a programmable classical initial state and thus enables sequential classical-quantum processing. Specifically:

\begin{figure}
\centering
\includegraphics[width=\linewidth]{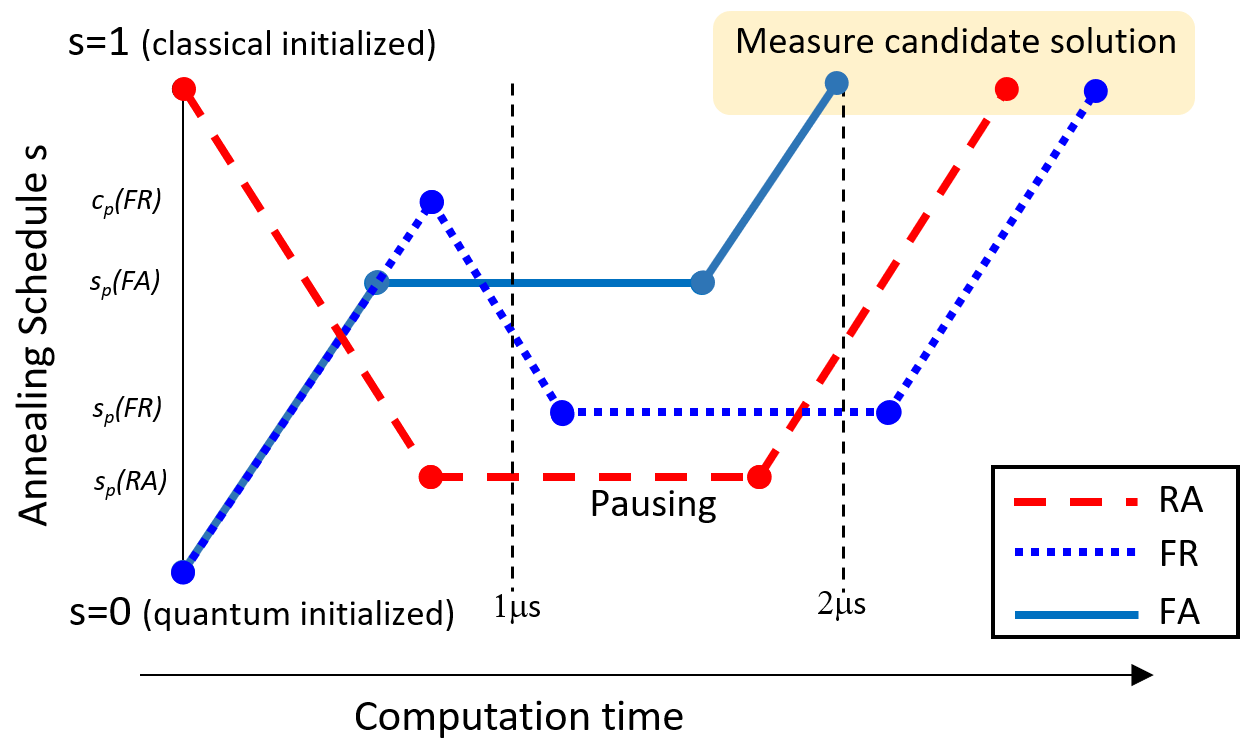}
\caption{The three flavors of quantum annealing (FA, RA and FR) investigated in this study. The annealing schedule $s$ determines the inverse strength of the signal that uses quantum effects to compute. If $s=0$, the quantum annealer realizes a fully quantum state that would return a random bitstring if measured. For intermediate $0\le s\le1$, the annealer is optimizing exploiting quantum fluctuations to explore the feasible solutions. At $s=1$ quantum fluctuations are suppressed and the system can be treated as a classical memory register with a stored result.} 
\label{f:annealing}
\end{figure}

\begin{figure*}
\centering
\includegraphics[width=\linewidth]{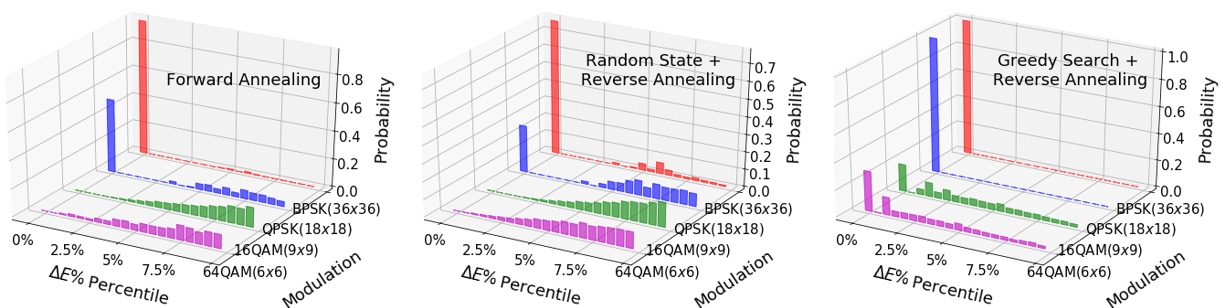}
\caption{Average distribution of cost function value percentile out of 200,000-600,000 anneal samples of 20 instances of 36-variable decoding problems for different modulations and \textbf{algorithms}: \emph{(Left)} forward annealing or QuAMax~\cite{kim2019leveraging}, \emph{(Center)}: reverse annealing starting at a randomly picked initial state, \emph{(Right)} reverse annealing starting at the result state of greedy search (hybrid processing with the simplest classical solver).}
\label{f:cdf_reverse}
\end{figure*}

\parahead{(1) Classical Module: initial greedy search.} Initially, GS solves the QUBO  with a candidate solution determined by greedy descent~\cite{ReverseVenturelli}.The bits are sorted in ascending order by the magnitude of $|\frac{1}{2}Q_{ii} + \frac{1}{4} \sum^{i-1}_{k=1} Q_{ki} + \frac{1}{4} \sum^{N}_{k=i+1} Q_{ik}|$.\footnote{In other words, the bits are sorted by the absolute magnitude of matrix's diagonal elements in the Ising model.}
The first bit is assigned $q_i = 0$ if the corresponding magnitude is positive and $1$ otherwise. Then the procedure is iterated recursively on the remaining variables by assigning the value that minimize the energy of the QUBO form considering only the variables that are set, until all variables are determined. As expected, the greedy solution is often not the global optimum, but it is a good initial guess that requires nearly negligible computation time and resources. By choosing the simplest classical module, we focus on viability of hybrid design, excluding the discussion on design challenges introduced in Section~\ref{s:design}.

\parahead{(2) Quantum Module: reverse annealing.} RA is a refined local quantum annealing starting from a known initial state, which has shown improved performance over FA for many different applications~\cite{ReverseVenturelli,ottaviani2018low,terry2020quantum}. With reference to Figure \ref{f:annealing}, in our prototype, RA is initialized with the solution classical state of GS ($s=1$) and starts annealing backward ($R$) up to $s_p$ ($0.0\leq s_p\leq 1.0$), pausing ($P$) for $t_p$ microseconds\footnote{It has been shown that the annealing pause brings out improvements for FA \cite{kim2019leveraging,DWPauseMarshall,izquierdo2020ferromagnetically} and for RA \cite{ReverseVenturelli}.}, before switching to forward ($F$). 
The comparison between the RA and the FA programming steps can be stated in terms of the schedule steps [time (us), $s$ ($0.0-1.0$)] follows:

\begin{itemize}
    \item Forward Annealing (FA):  

$[0.0,0.0] \xrightarrow{F} [s_p,s_p] \xrightarrow{P} [s_p+t_p, s_p] \xrightarrow{F} [t_a+t_p, 1.0]$

    \item Reverse Annealing (RA): 

$[0.0,1.0] \xrightarrow{R} [1.0-s_p, s_p] \xrightarrow{P} [1.0-s_p+t_p, s_p] \xrightarrow{F} [2(1.0-s_p)+t_p, 1.0]$,

\end{itemize}

\parabreak  where $t_a$ is \emph{anneal time}, $t_p$ is anneal \emph{pausing time}, and the mid-point phase $s_p$ denotes \emph{pause location} for FA and \emph{switch+pause location} for RA. These are all parameters that need to be optimized. While total duration of FA depends on anneal time $t_a$, RA total duration depends on switch and pause location $s_p$.  



\parahead{Forward reverse annealing.} In addition to FA, we test a newly-developed fully-quantum forward reverse annealing (FR) as another comparison scheme. The combination of FA+RA (\emph{i.e.,} FA$\rightarrow$ solution state $\rightarrow$RA) has been tested before to solve Low-Rank Matrix Factorization~\cite{ottaviani2018low}, in two steps, but in our implementation we program the annealer to execute a forward reverse annealing in a single step, where RA initial state is determined by the FA procedure without doing a measurement, initializing the reverse annealing with the state at s=$c_p$. FR's anneal schedule is the following:

\begin{itemize}
    \item Forward Reverse Annealing (FR): 

$[0.0,0.0] \xrightarrow{F} [c_p, c_p] \xrightarrow{R} [2c_p-s_p, s_p] \xrightarrow{P} [2c_p-s_p+t_p, s_p] \xrightarrow{F} [2c_p-2s_p+t_p+t_a, 1.0]$.
\end{itemize}

\subsection{Implementation}
\label{s:implementation}
We implement our prototype on the D-Wave 2000Q, a state-of-the-art analog-based quantum device to explore the possibility of improvement over pure quantum processing.  
We set the anneal time to be $t_a=1\mu s$ for FA (the minimum compute time allowed by the hardware) and the pausing time $t_p=1\mu s$ for FA, FR, and RA, consistently to the guidance in the literature for best performance. For $s_p$ and $c_p$, we set its range from 0.25--0.99 (in steps of 0.04). 
We collect statistically enough anneal samples per setting ($N_s$ at least 10,000) to analyze the performance. 
We synthesize 10-20 (QUBO) instances of random MIMO detection for various user numbers and modulations (BPSK, QPSK, 16-QAM, and 64-QAM) with unit gain signal and unit gain wireless channel with random phase. The method of reducing MIMO detection into QUBO form based on mapping rule between QUBO variables and wireless symbols was introduced in Ref.~\cite{kim2019leveraging} and we apply the same mapping.
In the experiments, we exclude the wireless noise (AWGN). For FA, we use data from the paper~\cite{kim2019leveraging}, and different algorithms are evaluated for the same QUBO instances as FA.

\subsection{Experimental Results}
\label{s:evaluation}



A fundamental metric to evaluate heuristics\hyp{}based solvers is the distribution of QUBO value of the samples.  We define the quality of solution as the percentile of solution cost $E_s$ compared to the best possible solution whose cost is $E_g$: 
$$\Delta E\%= 100 \cdot\left[( \abs{E_g} - \abs{E_s})/ \abs{E_g}\right].$$
$\Delta E\%=0\%$ indicates that the global optimum has been found. Figure~\ref{f:cdf_reverse} shows the empirical distribution  $\Delta E\%$ out of all anneal samples of a 36-variable MIMO detection solved by FA and RA with median best parameter setting. The lower $\Delta E\%$ means the closer gap between $E_s$ and $E_g$, which implies closer Euclidean distance to the optimal detection symbol. 
If the state that initializes RA is randomly selected  (Figure~\ref{f:cdf_reverse}, \emph{center}) then the method works worse than FA (Figure~\ref{f:cdf_reverse}, \emph{left}), skewing the distribution towards low quality solutions. The distribution obtained by running RA after the GS initialization (Figure~\ref{f:cdf_reverse}, \emph{right}) shows the most promise and hence it is used for our classical-quantum hybrid prototype design. 

\begin{figure}
\centering
\includegraphics[width=0.9\linewidth]{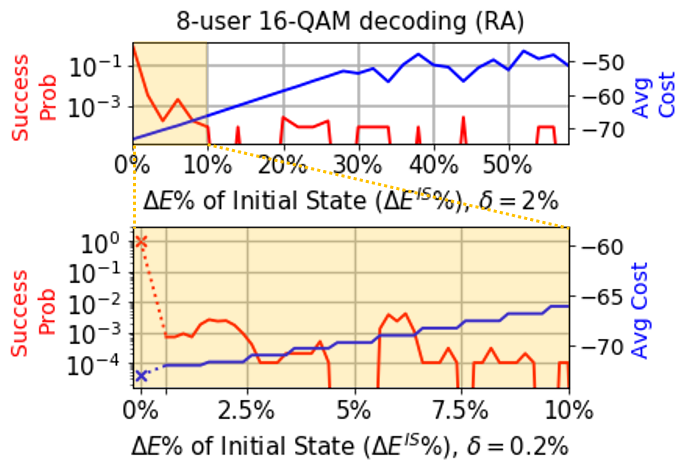}
\caption{Expectation value of the cost function and success probability out of RA samples for a 8-user 16-QAM decoding instance across different $\Delta E^{IS}\%$. Shaded parts are a close up of the results for $\Delta E^{IS}\%$ below 10\% (No initial candidate achieved less than 0.4\%).}
\label{f:reverse_initial_state}
\end{figure}

\parahead{Impact of the quality of the initial state.} We observe that most of the solutions returned by GS are already scoring approximately $\Delta E\% \leq 10\%$. In order to design our hybrid architecture, we study the correlation between the quality of RA's initial state ($\Delta E^{IS}\%$) and the overall solver performance after the RA. 
We obtain sample states of various $\Delta E^{IS}\%$  using over 750,000 samples. Figure~\ref{f:reverse_initial_state} shows the success probability of finding the optimal solution  and average cost computed out of RA samples in one typical 8-user 16-QAM detection instance as a function of $\Delta E^{IS}\%$ (binned in steps of $\delta = 2\%$). 
We observe that, unsurprisingly, the probability of success and the expectation value for the cost function is generally better if the $\Delta E^{IS}\%$ is low.



\begin{figure}
\centering
\includegraphics[width=0.95\linewidth]{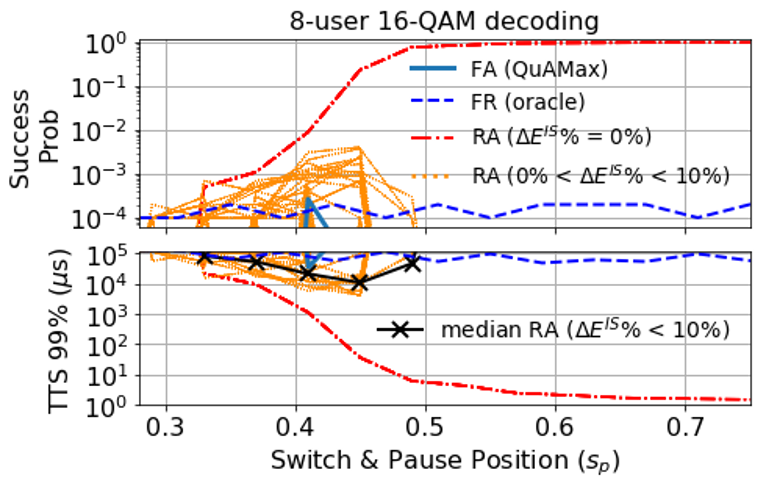}
\caption{Success probability and TTS of RA compared against FA and FR for a 8-user 16-QAM decoding instance, initialized with different methods and candidate solutions of various quality $(\Delta E^{IS}\%)$. The performance is reported as a function of the parameter $(s_p)$.}
\label{f:reverse_pause}
\end{figure}

\parahead{Impact of switch and pause location ($s_p$).} The performance is dependent on the parameter "switch and pause location" $s_p$ ($0 \leq s_p \leq 1$). $s_p$ should not be too close to $1$, since quantum fluctuations require to be strong enough to perturb the initialized state. At the same time, $s_p$ cannot be too close to $0$, since the information related to the initial state would be wiped out by too strong fluctuations, countering all possible advantages of RA as a "refined local search". Recall that the choice of $s_p$ affects the total duration of the computation. For our benchmark, we use a commonly-accepted metric for performance of quantum heuristics, \emph{time-to-solution} (TTS)~\cite{ronnow2014defining}, which indicates the median required time (us) to find the global optimum with \emph{target confidence} $C_t$\%:
\begin{small}
\begin{equation}\label{eq:n-rep}
    TTS(C_t\%) = duration\cdot \frac{\log(1-C_t/100)}{\log(1 - p^\star)}, 
\end{equation}
\end{small}
where $p^\star$ is the probability to find the ground-state during a single execution of the solver (FA, FR, and RA).
We plot their $p^\star$ and TTS at $C_t=99\%$ across different $s_p$ in Figure~\ref{f:reverse_pause}. The red dashed line denotes the RA run with the ground state as a RA initial state ($\Delta E^{IS}\% = 0\%$) and each dotted yellow line denotes RA runs initialized with various initial states of $0 < \Delta E^{IS}\% < 10\%$ ($\delta=0.2\%$). The initial states featuring $\Delta E^{IS}\% \geq 10\%$ are out of our consideration since they are very rarely generated. While FA cannot find the global optimum for the pause locations tested other than $s_p = 0.41$, RA is successful for $s_p$ chosen in the interval 0.33 to 0.49. In the case of FR, $p^\star$ (and thus TTS) result is reported for the best found $c_p$ out of an exhaustive search (the "oracle" scheme). The performance of the method for this instance happens to be worse than FA and RA despite the $c_p$ optimization. 

\section{Conclusion}
\label{s:concl}

In this paper, we outlined the landscape of hybrid classical-quantum processing in wireless networks. We share lessons that we learn from possible designs and explore the feasibility of a hybrid design enabled by reverse annealing. Our experimental results on a real hardware prototype show RA achieves improved performance comparing against two fully quantum solvers, even with the simplest classical solver which frequently fails to find the global optimum. Of course, the current state of maturity of the quantum annealing technology, as well as of other quantum-hardware, still forbids real-world deployment at scale due to high costs and system integration overheads that make the use impractical~\cite{kim2019leveraging}. Further, the results reported are mostly illustrative since they relate to performance on a single typical problem instance. Nevertheless, the speedup observed is a promising indication that a full\hyp{}scale bench\hyp{}marking of hybrid reverse annealing solvers is worthwhile to advance the state of art.

One important next step consists in exploring RA\hyp{}based hybrid designs with application-specific solvers.
The combination of application-specific classical solvers and RA is very likely to improve over the GS initialization. Classical approximate solvers for possible combinations with RA include (but are not limited to) linear solvers and tree search-based solvers. Linear solvers (\emph{e.g.,} zero-forcing) can likely achieve better initialization quality $\Delta E^{IS}\%$ than GS, requiring matrix inversion to nullify the wireless channel effect and thus slightly longer compute time, but their process cannot be parallelized. Tree-based solvers (\emph{e.g.,} FCSD~\cite{FCSD} and K-best SD~\cite{Guo06}) have tunable complexity, enabling parallelism, which could provide some control over $\Delta E^{IS}\%$. However, this flexibility comes at the expense of more complex hybrid designs.

\section*{Acknowledgements}
We thank the anonymous reviewers of this paper for their extensive technical feedback, which has enabled us to significantly improve the work. 
We also thank the NASA Quantum AI Laboratory (QuAIL), and the Princeton Advanced Wireless Systems (PAWS) Group for useful discussions. 
This research is supported by National Science Foundation (NSF) Award CNS\hyp{}1824357 and CNS\hyp{}1824470, and an award from the Princeton University School of Engineering and Applied Science Innovation Fund. Minsung Kim was also supported by the USRA Feynman Quantum Academy funded by the NAMS R\&D Student Program at NASA Ames Research Center (FA8750-19-3-6101).


\bibliographystyle{ACM-Reference-Format}
\bibliography{sample-base}

\end{document}